\documentclass[aps,pra,twocolumn,showpacs,groupedaddress]{revtex4-1}   % for review and submission with two columns
\usepackage{hyperref}  % needed for live references
\usepackage{graphicx}  % needed for figures
\usepackage{dcolumn}   % needed for some tables
\usepackage{bm}        % for math
\usepackage{amssymb}   % for math
\usepackage{verbatim}  % for comment
\usepackage{epstopdf}
\usepackage{color}
\newcommand{\ket}[1]{\ensuremath{\left| #1 \right\rangle}}
\newcommand{\bra}[1]{\ensuremath{\left\langle #1 \right|}}

\newcommand{\ave}[1]{\ensuremath{\left\langle #1 \right\rangle}}
\newcommand{\mele}[3]{\ensuremath{\left\langle #1 \right|#2\left| #3
\right\rangle}}
\newcommand{\dyad}[2]{{\ket{#1}\!\!\bra{#2}}}

\newcommand{\beq}{\begin{equation}}
\newcommand{\eeq}{\end{equation}}
\newcommand{\bea}{\begin{eqnarray}}
\newcommand{\eea}{\end{eqnarray}}

\newcommand{\eq}[1]{{(\ref{#1})}}
\newcommand{\commentout}[1]{{}}

\newcommand{\half}{{\hbox{$\frac{1}{2}$}}}

\definecolor{red}{rgb}{1,0,0}

\begin{document}

\title{Ground state of the double-well condensate for quantum metrology}

\author{Juha Javanainen}
\affiliation{Department of Physics, University of Connecticut, Storrs, Connecticut 06269-3046}

\author{Han Chen}
\affiliation{Department of Physics, University of Connecticut, Storrs, Connecticut 06269-3046}

\begin{abstract}
We discuss theoretically the ground state of a Bose-Einstein condensate with attractive atom-atom interactions in a double-well trap as a starting point of Heisenberg-limited atom interferometry. The dimensionless parameter governing the quality of the ground state for this purpose is identified. The near-degeneracy between the ground state and the first excited state severely curtails the prospects of the thermally prepared ground state in quantum metrology.
\end{abstract}
\maketitle
\section{Introduction}
The precision of all measurements is ultimately constrained by quantum mechanics~\cite{HEL76,BRA94,BRA96}, but quantum mechanics also brings opportunities to improve the precision beyond what might appear feasible on the basis of classical arguments. For instance, the precision of interferometry could be improved by using various special quantum input states~\cite{HOL93,BOL96}, such as the ``Schr\"odinger cat'' (SC) state~\cite{BOL96}. Thereby the error might scale with the number of particles $N$ (say, photons) according to the Heisenberg limit $1/N$, instead of the classical shot-noise limit $1/\sqrt{N}$~\cite{LEE02}.

While we are not aware of any practical quantum limited interferometry carried out with a Bose-Einstein condensate in a double-well trap, the analogy with optical interferometry suggests that Heisenberg-limited measurement of the energy difference of the atoms between the two sides of the trap is possible in principle, and as such could be used for, say, measurements of small forces acting on the atoms. The novel feature compared with optical interferometry is readily present modifications of the measurement precision owing to atom-atom interactions~\cite{GRO11,JAV12NLI}. This paper, however, works another aspect of atom-atom interactions noted in Ref.~\cite{JAV12NLI}: If the condensate atoms have a strong attractive interaction among themselves, the ground state is close to a SC state~\cite{ZOL98}, and as such might appear to be a good starting point for Heisenberg-limited interferometry. 

We study the use of the ground state in interferometry both analytically and numerically in a scheme in which the parity of the atom number in one or the other of the potential wells is ultimately measured~\cite{BOL96}. Our main result is to identify the dimensionless parameter that governs the quality of the ground state for interferometry. Given the requirements for Heisenberg-limited interferometry, we may then ask when they can be fulfilled. The issue here is that the ground state and the lowest-energy excited state become degenerate in the limit of strong atom-atom interactions, and thermal preparation will give a mixture of the two states that is useless in interferometry. Our findings are somewhat discouraging: The problem with the near-degeneracy gets  exponentially worse with the atom number, and is likely to negate much of the potential advantage from near-SC states in high-precision interferometry.

\section{Metrology with Schr\"odinger cat}
We start with the usual two-mode model, the Bose-Hubbard model with two sites, for the double-well potential~\cite{MIL97}, writing the Hamiltonian
\bea
H &=& -J(b^\dagger_\ell b_r + b^\dagger_r b_\ell)  + U(b^\dagger_\ell b^\dagger_\ell b_\ell b_\ell + b^\dagger_r b^\dagger_r b_r b_r)\nonumber\\
&&+\half\epsilon(b^\dagger_\ell b_\ell - b^\dagger_r b_r)\,.
\label{HAM}
\eea
Here $b_\ell$ and $b_r$ are the annihilation operators for bosons on the ``left'' and ``right'' sides of the double-well trap, $J$ is a frequency that characterizes tunneling between the two wells, $U$ is a measure of the on-site atom-atom interactions, and $\epsilon$ characterizes the energy difference between the potential wells. We set $\hbar=1$, and regard energy and (angular) frequency as interchangeable. Here we only consider attractive atom-atom interactions, or more precisely, $U\le0$. We study the model~\eq{HAM} both analytically and numerically. For the latter purpose we have a collection of analysis objects written in $C++$, materially the same that were employed in Ref.~\cite{JAV13}. The nontrivial linear algebra is done with LAPACK.

It is expedient to transform the Hamiltonian to an alternative  Schwinger representation. We define
\bea
S_x = \half(b^\dagger_\ell b_r + b^\dagger_r b_\ell),\nonumber\\
S_y = \hbox{$\frac{1}{2i}$} (b^\dagger_\ell b_r - b^\dagger_r b_\ell),\nonumber\\
S_z = \half(b^\dagger_\ell b_\ell - b^\dagger_r b_r)\,.
\eea
These operators obey the SU(2) or angular momentum algebra (with $\hbar=1$), $S=N/2$ acting as the quantum number for the total angular momentum. Except for a constant of the motion that has no effect on either dynamics or thermodynamics, in the angular momentum representation the Hamiltonian reads
\beq
H = -2JS_x + 2 U S_z^2+ \epsilon S_z\,.
\eeq
Perhaps surprisingly, in order to keep both the standard many-body formalism and the standard angular-momentum methods intact, there are some nonobvious relations between the phases of the eigenstates of the components of angular momentum. For instance, if we declare that the state with all $N$ atoms in the left trap $\ket{n_\ell=N,n_r=0}\equiv\ket{N,0}$ is the eigenstate of $S_z$ with the eigenvalue $m_z=N/2$, then the normalized eigenvector corresponding to the eigenvalue $m_z=-N/2$ is uniquely set to be $\ket{0,N}$. Consult the Appendix for more details on the use of angular momentum algebra in this work.

As has been known for a while~\cite{ZOL98}, in the case of strong attractive interactions with $|U|$ asymptotically large, and for $\epsilon=0$, the ground state of the model~\eq{HAM} is the Schr\"odinger cat (SC)
\beq
\ket{S} = \frac{1}{\sqrt2}(\ket{N,0} + \ket{0,N})\,,
\label{SCHP}
\eeq 
a superposition of the states with all atoms in either the left or the right well. The first excited state in this limit is also a SC,
\beq
\ket{A} = \frac{1}{\sqrt2}(\ket{N,0} - \ket{0,N})\,.
\label{SCHM}
\eeq

On the other hand, measuring the energy difference $\epsilon$ in the Hamiltonian~\eq{HAM} is a variant of the prototypical interferometry problem. The generic scheme is to prepare the system in an initial state, let it evolve in a manner that depends on the value of the parameter to be measured, and finally infer the value from some measurement.
Arguments based on quantum Fisher information~\cite{HEL76,BRA94,BRA96} tell us that in linear interferometer in which the coupling Hamiltonian of the probe to the quantity to be measured $\theta$ is of the form $H'=-\theta \hat N$, after optimizing over both all possible measurements and all possible initial states, the smallest achievable standard deviation the measured $\theta$ equals the Heisenberg limit $1/N$, and that the optimal precision may in principle be reached starting from a SC state such as~\eq{SCHP} or \eq{SCHM}. This analysis even comes with a prescription of the measurement to be made to reach the best possible precision, but with no guarantee that the measurement can be carried out in practice. Interestingly, though, in interferometry a realizable (in principle) measurement scheme exists that reaches the optimal precision~\cite{BOL96}. In this paper we address such ``parity measurements.'' We always assume that, by manipulating the strengths of the lattice lasers and by making use of Feshbach resonances, the parameters of the Hamiltonian may be controlled at will. How things work out when the control is not complete, in particular, when atom-atom interactions are present during the interferometry, is discussed, e.g., in Refs.~\cite{GRO11} and~\cite{JAV12NLI}.

Let us now assume that one sets the system to some initial states $\ket{\psi_0}$; ideally the SC state~\eq{SCHP}, but for the moment we allow an arbitrary input state. The parameter $\epsilon$ is then allowed to settle to a value that is to be measured. For simplicity let us stipulate that the other parameters $J$ and $U$ are zero at this stage.  Thus, the time evolution operator is $U(t) = e^{-iS_z \epsilon t}$. We write the evolution phase in the form $\theta=\epsilon t$. This is ultimately the parameter to be measured, the smallest possible standard deviation being $\sigma_\theta=1/N$. The phase gets imprinted on the system with the evolution operator $U(\theta)=e^{-iS_z \theta}$, so that the state becomes $\ket{\psi_1}=U(\theta)\ket{\psi_0}$.

Next, in the optimal scheme the energy difference to be measured is at least functionally turned off and a suitable strength of tunneling is applied for a preset time with $4Jt= \pi$ so that the time evolution is the unitary transformation $R = e^{-i (\pi/2) S_x}$, rotation of the initial state about the $x$ axis by the angle $\pi/2$. After this ``beam splitter'' operation the state is $\ket{\psi_2}= R\ket{\psi_1}$. Finally the number of the atoms in the traps is measured. The decisive quantity~\cite{BOL96} is the parity of the atom number in, say the left trap, $\cal P$; $(-1)^{n_\ell}$ for $n_\ell$ atoms. The corresponding Hermitian operator is
\beq
{\cal P} = \sum_{n=0}^N (-1)^n \dyad{n,N-n}{n,N-n}\,.
\eeq
The eigenvalues of $\cal P$ are $\pm 1$, and $\cal P$ is also its own inverse,
\beq
{\cal P}^2 = 1\,.
\eeq

Suppose the initial state, in fact, is the  SC~\eq{SCHP},
\beq
\ket{\psi_0} = \frac{1}{\sqrt2}(\ket{N/2}_z+\ket{-N/2}_z)\,,
\eeq
where we have emphasized that the state $\ket{N,0}$ ($\ket{0,N}$) is also the eigenstate of the $z$ component of the fictitious spin with the eigenvalue $N/2$ ($-N/2$). The state after the phase to be found has been imprinted on the system is
\beq
\ket{\psi_1} = \frac{1}{\sqrt2}(e^{-iN\theta/2}\ket{N/2}_z+e^{iN\theta/2}\ket{-N/2}_z)\,.
\eeq
The rotation operator converts the eigenstates $\ket{\pm N/2}_z$ to the corresponding eigenstates $\ket{\mp N/2}_y$ of the $y$ component of the angular momentum, so that before the final measurement the state is
\beq
\ket{\psi_2} = \frac{1}{\sqrt2}(e^{-iN\theta/2}\ket{-N/2}_y+e^{iN\theta/2}\ket{N/2}_y)\,.
\eeq
Finally, the parity operator $\cal P$ is measured in this state.

To flesh out this description we first note that the relevant eigenstates of $S_y$ are
\bea
\ket{N/2}_y &=&\sum_{n=0}^N \left[\frac{1}{2} {N\choose n}\right]^{1/2}(-i)^n \ket{n,N-n},
\label{SYP}\\
\ket{-N/2}_y &=& (-i)^N \sum_{n=0}^N \left[\frac{1}{2} {N\choose n}\right]^{1/2}i^n \ket{n,N-n}\,.\,
\label{SYM}
\eea
These are easily verified to be eigenstates of $S_y$ as claimed, and in fact they also have the proper phases consistent with the rotations of $\ket{-N/2}_z$ and $\ket{N/2}_z$ about the $x$ axis by the angle $\pi/2$.

Now, the parity operator $\cal P$ also has the property that
\beq
{\cal P} \ket{-N/2}_y =(- i)^N \ket{N/2}_y\,.
\label{YFLIP}
\eeq
By the time of the final measurement the state therefore is
\beq
\ket{\psi_2} = \frac{1}{\sqrt2} (e^{iN\theta/2} \ket{N/2}_y  + (-i)^Ne^{-iN\theta/2}{\cal P}  \ket{N/2}_y)\,.
\eeq
This gives the expectation values of parity and of the square of parity, and the standard deviation of parity, in the form
\bea
{\cal P}(\theta)&=&\ave{\cal P} = \mele{\psi_2}{{\cal P}}{\psi_2} = 
\cos[N(\theta+\pi/2)],\commentout{
\left\{
\begin{array}{ll}
(-1)^{N/2} \cos N\theta,  &N\,\,{\rm even},  \\
(-1)^{(N+1)/2} \sin N\theta,  &N\,\,{\rm odd}, \\
\end{array} 
\right.}\label{ORIGINALP}\\
\ave{{\cal P}^2} &=& 1,\\
\sigma_{\cal P}(\theta) &=& \sqrt{\ave{{\cal P}^2}-\ave{{\cal P}}^2}=
|\sin[N(\theta+\pi/2)]|\commentout{
 \left\{
\begin{array}{ll}
|\sin N\theta|, &N\,\,{\rm even}, \\
|\cos N\theta| ,  &N\,\,{\rm odd}. \\
\end{array}
\right.}.
\eea
Finally, the error propagation formula reads
\beq
\sigma_\theta = \left |\frac{\sigma_{\cal P}(\theta)}{{\cal P}'(\theta)} \right|,
\label{ERRPROP}
\eeq
where ${\cal P}'(\theta)$ is the derivative of the expectation value of parity ${\cal P}(\theta)$ with respect to the angle $\theta$. The error propagation formula shows that the standard deviation for the inference of the angle $\theta$ is $\sigma_\theta=1/N$, the standard Heisenberg limit~\cite{BOL96}, and the best one can do in linear interferometry.

For an odd number of atoms the parity starts from zero at $\theta=0$, as in ${\cal P}(\theta)\propto\sin N\theta$, while for an even $N$ we have a extremal value of parity $\pm 1$ at $\theta=0$; ${\cal P}(\theta)\propto\cos N\theta$. The parity obviously oscillates with the phase to be measured $\theta$, but irrespective of atom number $N$ there always is an extremum with ${\cal P}(\theta)=\pm1$ at $\theta=\pi/2$ and $3\pi/2$. Since ${\cal P}(\theta)$ has zero derivative at an extremum, the observable ${\cal P}(\theta)$ does not depend on the value to be measured $\theta$ at all and one expects a poor performance of a measurement, however at these same points the standard deviation of parity also vanishes and one finds the Heisenberg limit $1/N$ for all $\theta$. How useful the extrema are in actual measurements depends on the details of the experiment. We will not attempt to resolve such technical issues, but simply take the results of our analysis at face value.

In actual laboratory practice things would not work out quite so readily even if the system parameters such as $N$, $J$ and $U$ could be controlled to an arbitrary precision. For one thing, an initial Schr\"odinger cat state is not easy to prepare or preserve. Second, miscounting just one single  particle plays havoc with the parity. For the present purposes we assume that control of atom number and detection efficiency are not  problems, and analyze the effect of the initial state on ideal measurements of parity. In fact, we believe that detection efficiency will be less of an issue with atoms than with the ephemeral photons; techniques capable of single-atom resolution out of a hundred~\cite{ZHA12-PRL} and a thousand~\cite{HUM13} atoms have already been demonstrated. The calculations were done analytically for the initial Schr\"odinger cat, and even in the general case the derivative of the expectation value ${\cal P}'(\theta)$ may be found directly as an expectation value of a certain operator with no need to find a derivative numerically. We outline this argument in the Appendix.

\section{Metrology in the ground state}
\subsection{Precision in the ground state}
Assume now that the system starts in the ground state of the double-well system in the limit when the ground state is almost but not quite the SC~\eq{SCHP}. This means that atom-atom interactions $\propto U$ are dominant, and for the moment we think of tunneling $\propto J$ as a perturbation.

Assuming no tunneling at all, $\ket{\pm N/2}_z$ are both degenerate lowest-energy states of the system with the energies $N^2U/2$. The perturbation first couples in the states $\ket{\pm (N/2-1)}_z$ with the energies $2(N/2-1)^2U$. To the order $J^1$ we therefore have, say, the state
\beq
\ket{+} = \ket{N/2}_z - \xi S_x  \ket{N/2}_z,\quad \xi = \frac{J}{(N-1)U}\,,
\eeq
and analogously the state $\ket{-}$ corresponding to the unperturbed state $\ket{-N/2}_z$.
These expressions are singular for $N=1$ because then there is no atom-atom interaction to start with. Henceforth we take $N>1$.

The perturbation cannot split the degeneracy of the states $\ket{\pm N/2}_z$ in the order $J^1$, so we need to construct the counterparts of the SC states $\ket S$ and $\ket A$ manually.  All told, we have the ground state, and the initial state for the interferometry, to the leading order in perturbation theory in the form
\beq
\ket{\tilde\psi_0} = (1-\xi S_x)\ket{S}\,.
\eeq
Next, applying the evolution operator $U(\theta)$, we find that after the energy difference to be measured is imprinted on the system, the state is
\beq
\ket{\tilde\psi_1}\! =\!\ket{\psi_1}- \frac{\xi}{\sqrt{2}}S_x\! \left[\!  e^{-i\theta(\frac{N}{2}-1)}\! \ket{N/2}_z
\! \!+\! e^{i\theta(\frac{N}{2}-1)}\! \ket{-N/2}_z\!  \right].
\eeq
Here and below the states $\ket{\psi_1}$, etc., without the tildes are the states following from the ideal initial state $\ket S$. After the rotation about the $x$ axis, the state is
\bea
\ket{\tilde\psi_2}\!  &=&\!  \ket{\psi_2} - \xi \ket{\Psi_2};\nonumber\\
\ket{\Psi_2}\!  &=&\!\!   \frac{1}{\sqrt{2}}S_x\! \! \left[\!  e^{i\theta(\frac{N}{2}-1)}\! \ket{N/2}_y\!\!  +\! 
e^{-i\theta(\frac{N}{2}-1)}\! \ket{-N/2}_y\! \right].\nonumber\\
\eea
Finally, the expectation value of parity is
\bea
\mele{\tilde\psi_2}{{\cal P}}{\tilde\psi_2} &=& {\cal P}(\theta)\label{ZERO}\\
&&-\xi\left( \mele{\psi_2}{{\cal P}}{\Psi_2} + \mele{\Psi_2}{{\cal P}}{\psi_2}\right)\label{ONE}\\
&&+\xi^2 \mele{\Psi_2}{{\cal P}}{\Psi_2}\label{TWO}\,.
\eea

The zeroth-order contribution~\eq{ZERO} is the same as before, Eq.~\eq{ORIGINALP}. The first-order contribution~\eq{ONE} consists of terms of the form
\bea
&&\mele{\psi_2}{{\cal P}}{\Psi_2} =
\frac{1}{2}\left[e^{i\theta(N-1)}{}_y\!\mele{-N/2}{{\cal P}S_x}{N/2}_y\nonumber\right.\\
&&\,+e^{i\theta}{}_y\!\mele{-N/2}{{\cal P}S_x}{-N/2}_y+e^{-i\theta}{}_y\!\mele{N/2}{{\cal P}S_x}{N/2}_y\nonumber\\
&&\,+\left.e^{-i\theta(N-1)}{}_y\!\mele{N/2}{{\cal P}S_x}{-N/2}_y\right]\!\!.
\eea
But, by virtue of Eq.~\eq{YFLIP}, matrix elements of the type ${}_y\!\mele{-N/2}{{\cal P}S_x}{N/2}_y$ are proportional to ${}_y\!\mele{N/2}{S_x}{N/2}_y$ and equal to zero, being essentially expectation values of a component of the angular momentum in an eigenstate of an orthogonal component. On the other hand, matrix elements of the form $_y\!\mele{-N/2}{{\cal P}S_x}{-N/2}_y$ are proportional to $_y\!\mele{N/2}{S_x}{-N/2}_y$. These also equal to zero, since $S_x$ can have nonzero matrix elements between two eigenstates of an orthogonal component of the angular momentum $S_y$ only if the respective quantum numbers  $m_y$ differ exactly by one, and here $N>1$. All told, there is no first-order correction $\propto J$ to the expectation value of parity.

Let us also consider the second-order correction in Eq.~\eq{TWO}. At this point we are then not pursuing perturbation theory consistently to second order, but it turns out that we learn valid lessons from this exercise anyway. We have
\bea
\mele{\Psi_2}{{\cal P}}{\Psi_2} &=& \frac{1}{2}\left[
{}_y\!\mele{N/2}{S_x{\cal P}S_x}{N/2}_y\right.\nonumber\\
&&+{}_y\!\mele{-N/2}{S_x{\cal P}S_x}{-N/2}_y\nonumber\\
&&+ e^{i(N-2)\theta}{}_y\!\mele{-N/2}{S_x{\cal P}S_x}{N/2}_y\nonumber\\
&&+ \left.e^{-i(N-2)\theta}{}_y\!\mele{N/2}{S_x{\cal P}S_x}{-N/2}_y\right].\nonumber\\
\label{SECORDSUM}
\eea
To simplify, we first note that in the representation of the eigenstates of $S_z$ the operator $S_x$ is a sum of dyads of the form $\ket{m_z}\!{}_z{}_z\!\bra{m_z\pm 1}$ that behave under the parity operator as 
\beq
{\cal P}\ket{m}\!{}_z{}_z\!\bra{m\pm 1}{\cal P}= -\ket{m}\!{}_z{}_z\!\bra{m\pm 1},
\eeq
and so we have
\beq
{\cal P}S_x{\cal P} = -S_x;\, {\cal P}S_x = -S_x{\cal P}\,.
\eeq
By an argument similar to the one we used for the first-order term, the first two terms on the right-hand side of~\eq{SECORDSUM} are zero at least if $N>2$. On the other hand, from an exercise in angular momentum algebra we have
\bea
{}_y\!\mele{-N/2}{S_x{\cal P}S_x}{N/2}_y &=& - i^N {}_y\!\mele{N/2}{S_x^2}{N/2}_y\nonumber\\
&=&-\half i^N {}_y\!\mele{N/2}{S_x^2+S_z^2}{N/2}_y\nonumber\\
&=&-\half i^N {}_y\!\mele{N/2}{{\bf S}^2-S_y^2}{N/2}_y\nonumber\\
&=&-\frac{i^N N}{4}\,.
\eea
We therefore have the second-order result
\beq
\xi^2 \mele{\Psi_2}{{\cal P}}{\Psi_2} = -\frac{NJ^2}{4(N-1)^2U^2} \cos[(N-2)\theta+N\pi/2].
\label{PSECO}
\eeq

Equation~\eq{PSECO}  constitutes only part of the second-order perturbative correction. Starting from the full second-order perturbation theory state vector (see, e.g., Chapter 38, Problem 1 in Ref.~\cite{LAN85}) and using similar techniques as above, it turns out that the full result up to second order in $J/U$ and for $N>2$ is
\bea
{\cal P}(\theta) &=& \left( 1- \frac{NJ^2}{4(N-1)^2U^2}\right )\cos[N\theta+N\pi/2]\nonumber\\
&& - \frac{NJ^2}{4(N-1)^2U^2} \cos[(N-2)\theta+N\pi/2]\,.
\eea
Several qualitative features are evident. First, for $N\gg1$ the leading correction to the expectation value of parity from the fact that the ground state is not an exact SC is proportional to the dimensionless parameter
\beq
\chi=\frac{J^2}{NU^2}.
\label{CHIDEF}
\eeq
Second, this correction and the leading contribution both have extrema at $\theta=\pi/2$ and $3\pi/2$, so that these points remain extrema of ${\cal P}(\theta)$ in spite of the second-order correction. Also, for odd $N$, $\theta=0$, $\pi$ and $2\pi$ remain zeros ${\cal P}(\theta)$. On the other hand, the other zeros and extrema shift with the perturbation, so there are no absolute phase markers except at  integer multiples of $\pi/2$. Numerically, the observations about the special role of the multiples of $\pi/2$ hold true for arbitrary problem parameters, and even in a strengthened form: the parity ${\cal P}(\theta)$  always has the extremal value $+1$ or $-1$ at $\pi/2$ and $3\pi/2$.

\begin{figure}
\includegraphics[width=8.0cm]{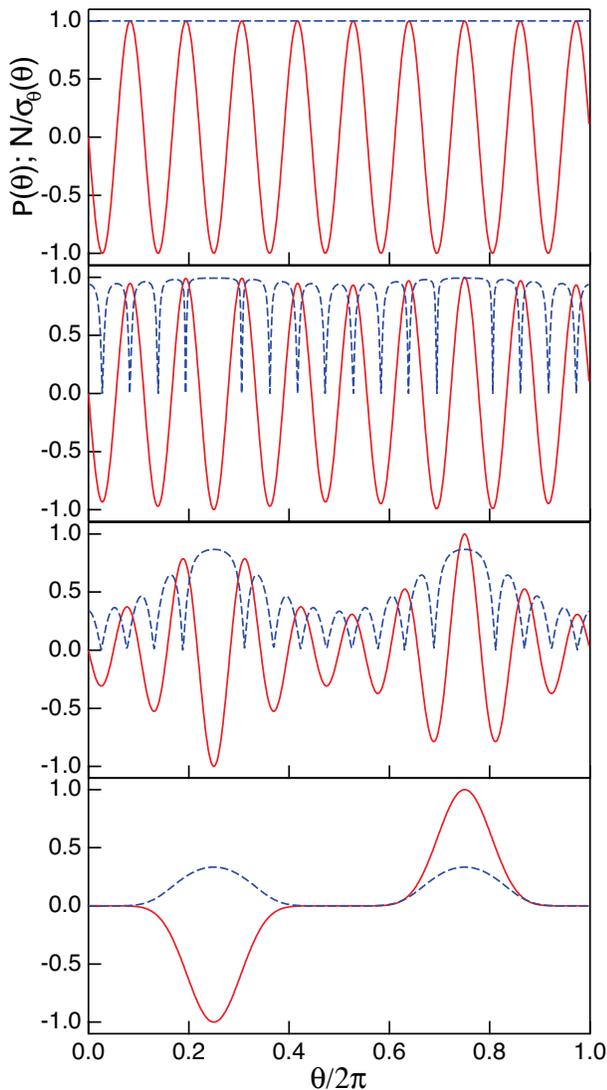}
\caption{Parity signal ${\cal P}$ (solid red line) and precision of the phase measurements $1/\sigma_\theta$ in units of the Heisenberg limit  (dashed blue line) as a function of the phase to be measured, starting from the ground state of the double-well system.  The fixed parameters are $N=9$, $J=1$, and the values of the atom-atom interaction strength are $U=-\infty$, $-1$, $-0.25$, and $0$ (top to bottom).
\label{FIG1}
}
\end{figure}

We illustrate with examples obtained by finding the ground state numerically. In Fig.~\ref{FIG1} we show both the parity signal ${\cal P}(\theta)$  (continuous red line) and the precision of parity measurements $1/\sigma_\theta$ from the error propagation formula~\eq{ERRPROP} (dashed blue line) as a function of the phase to be measured $\theta$. In the figure the precision is normalized to the Heisenberg limit $1/N$, so that the value $1$ represents the best possible measurement of the phase $\theta$. All of these examples are for $N=9$ atoms, we use the tunneling amplitude as the unit of frequencies by setting $J=1$, and the strength of the attractive atom-atom interaction decreases from top to bottom. The values, in fact, are $U=-\infty$ (pure SC), $U=-1$, $U=-0.25$, and $U=0$ (no atom-atom interaction at all). The corresponding values of the parameter $\chi$, Eq.~ \eq{CHIDEF}, are 0, 1/9, 16/9, and $\infty$.

The figure demonstrates the transition from very strong attractive atom-atom interactions to no interactions at all. Since $N=9$ is odd, the parity signal starts as zero at $\theta=0$, which means that the double-well system might then be particularly well suited for detection of small phases. As the interaction strength decreases, one first sees  in ${\cal P}(\theta)$ the beats associated with the simultaneous presence of the $\sin N\theta$ and $\sin[N(\theta-2)]$ signals. At $U=0$ the result is qualitatively different, but still displays the property that $P(\theta)=\pm 1$ at $\theta=\pi/2$ and $3\pi/2$. Regarding the measurement precision, unfortunately it peaks at $\theta=\pi/2$ and $3\pi/2$, and decreases  rather quickly with decreasing $|U|$ at $\theta=0$ that would be the natural operating point for measuring small angles $\theta$. In contrast, at $\theta=\pi/2$ the parity signal $P(\theta)$ is an extremum and a determination of $\theta$ from parity measurements is presumably maximally susceptible to errors not included in our model.

Overall, Fig.~\ref{FIG1} illustrates the following observation: If the ground state of the double-well trap can be prepared, it could be useful for near-Heisenberg limit resolution in the measurements when the atom-atom interactions are strong enough that the ground state is close to the SC. Quantitatively, this occurs when the value of the parameter $\chi=J^2/NU^2$ is at most on the order of one.

\subsection{Preparation of the ground state}
Barring secondary complications such as collapse of the condensates, outright failure of the two-mode model, etc., in the limits $U\rightarrow-\infty$ or  $N\rightarrow\infty$ the ground state of the double-well trap becomes doubly degenerate; the symmetric and antisymmetric SC states~\eq{SCHP} and~\eq{SCHM} are both ground states. In such a case the thermal density operator near zero temperature is
\beq
\rho = \half(\dyad{S}{S}+\dyad{A}{A}).
\label{THG}
\eeq
The parity signal  is the average of the signals for symmetric and antisymmetric SC states and is identically zero, so that no information about the angle $\theta$ is available.

\begin{figure}
\includegraphics[width=8.0cm]{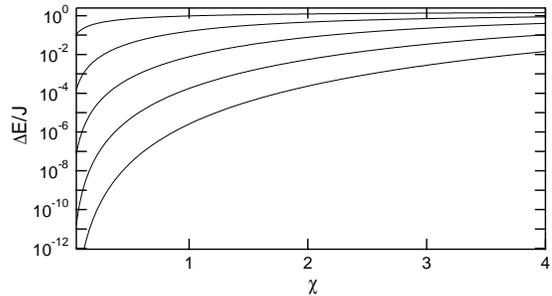}\\
\caption{Energy difference $\Delta E$ between the first excited state and the ground state of the double-well trap as a function of the parameter $\chi =  J^2/NU^2$ that governs the proximity of the ground state to the symmetric cat. The graphs are for atom numbers $N=3$, 6, 9, 12, and 15, from top to bottom}
\label{TWOAT}
\end{figure}

The problem here is that the energy difference between the two lowest-energy eigenstates $\Delta E$ tends to zero very quickly with increasing parameters $N$ or $|U|$, to the point that  general-purpose numerical eigensystem solvers fail to distinguish between the corresponding states. We demonstrate in Fig.~\ref{TWOAT}, where we plot $\Delta E$ as a function of the parameter $\chi=J^2/NU^2$ governing the quality of the ground state for the purposes of Heisenberg limited metrology; let us regard $\chi\sim1$ as the approximate upper limit. The various curves, from top to bottom, are for $N=3$, 6, 9, 12, and 15.

For a given $\chi$ and thus effectively for a given quality of the ground state, the energy difference between the first excited state and the ground state decreases basically exponentially with the atom number. This is a stern limitation on the preparation of the ground state. First, the temperature required for separation between the ground state and  the lowest excited state becomes unmanageably small. Second, a residual energy difference $\epsilon$ during the preparation comparable to $\Delta E$ is expected to eliminate the SC state. Third, the time required for preparation presumably balloons. Effectively, thermal preparation is possible only for small atom numbers, for which the difference between the Heisenberg-limit error $1/N$ and the shot-noise limit $1/\sqrt{N}$ is not particularly dramatic to begin with.

\section{Concluding remarks}
We have identified the dimensionless parameter $\chi= J^2/NU^2$ that characterizes the potential of the ground state of the double-well trap for Heisenberg-limit interferometry; the smaller $\chi$, the better. On the other hand, it then also becomes easy to quantify the serious problem that arises from the near-degeneracy of the ground state and the first excited state. In our view there is not much to gain in metrology from an attempt to prepare a SC state thermally.

There are also incidental limitations to the experiments in the form of the possible collapse of the condensate under the attractive interactions, and the two-mode model getting unreliable. Both of these eventualities are expected~\cite{DOD96, JAV99} when the interaction energy per atom in each potential well $\sim N|U|$ is comparable to the harmonic-oscillator energy $\omega$ of each of the two traps, $N|U|\sim\omega$. Combining with the estimate $\chi\sim1$ we find an approximate maximum number of atoms $N_m\sim (\omega/J)^2$. This can be made arbitrarily large in principle, by making $J$ small, but it would again be at the expense of a longer time scale in the experiment.

The thermal ground state~\eq{THG} can also be regarded as a mixture of SC states
\beq
\ket\varphi = \frac{1}{\sqrt 2}(\ket{N,0}+e^{i\varphi}\ket{0,N})\,
\eeq
over different angles $\varphi$, and all of the states $\ket\varphi$ could be used for Heisenberg limit metrology. A method of  determining the angle $\varphi$ (sufficiently) nondestructively would therefore facilitate high-precision metrology. One could think that, if there were a way to remove the atoms in, say, the state $\ket{A}$ while leaving the state $\ket{S}$ alone, half of the time the state $\ket{S}$ would be prepared; but again, we know of none. In view of the fundamental interest of the problem in general, it is no surprise that there have also been numerous discussions about active preparation of a SC state for a two-well system; Refs.~\cite{NOTEONEQUIV,RUO98CAT,STE98,GOR99,DAL00,RUO01SC,MIC03,HUA06,MAZ08,WAT10,LAP12} are a few. However, so far there apparently have been no experimental results.  We obviously have many interesting research questions here, but at this time few answers.
\section{Acknowledgments}
This work is supported in part by NSF, Grant No. PHY-0967644.

\appendix
\section{Some angular momentum algebra}
Take three components $S_x$, $S_y$ and $S_z$ of an angular momentum operator. They  have the commutator $[S_x,S_y]=iS_z$, and other commutators similarly from cyclic permutations. Angular momentum operators are generators of infinitesimal rotations. With this in mind, consider rotations about the $x$ axis as an example. We  define
\beq
S_z(\phi) = e^{-i S_x \phi} S_z e^{i S_x \phi},\quad S_y(\phi) = e^{-i S_x \phi} S_y e^{i S_x \phi}\,.
\label{APPROT}
\eeq
Using the commutators of the operators we immediately find
\beq
\frac{d}{d\phi} S_z(\phi) = - S_y(\phi),\quad \frac{d}{d\phi} S_y(\phi) =  S_z(\phi)\,.
\label{APPEQS}
\eeq
The simple expressions
\beq
S_z(\phi) = \cos\phi\, S_z - \sin\phi\, S_y,\, S_y(\phi) = \sin\phi\, S_z + \cos\phi\, S_y  
\label{EXPLROT}
\eeq
constitute a solution of Eqs.~\eq{APPEQS}, and  satisfy the initial conditions $S_z(0)=S_z$ and $S_y(0)=S_y$. Equations~\eq{EXPLROT} therefore give explicitly the operators defined in Eqs.~\eq{APPROT}.

On the other hand, let us define a $\half\pi$ rotated version of the eigenstate of the $z$ component of the angular momentum with the eigenvalue $m$ as
\beq
\ket\psi=e^{-i(\pi/2)S_x} \ket{m}_z.
\label{APPROTDEF}
\eeq
We then have
\beq
S_y \ket\psi = S_y e^{-i(\pi/2)S_x} \ket{m}_z\,,
\eeq
or using~\eq{APPROT} and~\eq{EXPLROT},
\bea
&&e^{i(\pi/2)S_x} S_y \ket\psi = e^{(i\pi/2)S_x} S_y e^{-i(\pi/2)S_x} \ket{m}_z\nonumber\\
&& = S_y(-\pi/2) \ket{m}_z = -S_z  \ket{m}_z = -m \ket{m}_z\,,
\eea
which immediately gives
\beq
S_y  \ket\psi = -m\, e^{-(i\pi/2)S_x} \ket{m}_z = -m \ket\psi\,.
\label{EVECEVAL}
\eeq
This result says that  $\ket\psi$ is an eigenvector of $S_y$ belonging to the eigenvalue $-m$. 

However, given $\ket{m}_z$,  Eq.~\eq{APPROTDEF} defines the vector $\ket\psi$ completely but the eigenvalue equation~\eq{EVECEVAL} only up to a phase factor. More generally, in order to be able to use the full range of angular momentum methods for the double-well potential, we need to figure out how to choose the relative phases of various states in the two-well representation so that the correct phase relations follow in the angular momentum representation.

Let us declare that the number state $\ket{N,0}\equiv\ket{n_\ell=N,n_r=0}$ is the eigenstate of $S_z$ with the eigenvalue $N/2$.  Since $e^{i\varphi}\ket{N,0}$ is likewise an eigenstate no matter what the value of the real phase $\varphi$ is, we are thereby setting a phase convention for the state $\ket{N/2}_z$, {\em and  also for all other angular momentum eigenstates}. As an example, states with the eigenvalues of the $z$ component for the angular momentum less that $N/2$ should be obtained by applying the lowering operator of the angular momentum $S_- = S_x-iS_y = b^\dagger_r b_\ell$. Given how the boson operators act on the number states, it is clear that the eigenstates of $S_z$ with decreasing eigenvalues $m=N/2,N/2-1,\ldots,-N/2$ have to be chosen as $\ket{N,0}$, $\ket{N-1,1}$,\ldots,\ket{0,N}. Likewise, the parity signal will depend on the relative phase of certain eigenstates of $S_y$, so that these eigenstates have to have proper phases. These phases are discussed implicitly in standard references on angular momentum, e.g.~\cite{EDM60}, but we have found the ones in~\eq{SYP} and~\eq{SYM} simply by carrying out the operations such in Eq.~\eq{APPROTDEF} numerically.

Finally, as noted in the main text, the derivative of the expectation value of the parity may be computed directly as an expectation value of a certain operator. We first note that
\bea
\ket{\psi_2'}&\equiv& \frac{d}{d\theta} \ket{\psi_2} = \frac{d}{d\theta} e^{-i(\pi/2) S_x} e^{-i\theta S_z}\ket{\psi_0}\nonumber\\
&=&  -i e^{-i(\pi/2) S_x} S_z  e^{-i\theta S_z}\ket{\psi_0}.
\eea
But, by combining Eqs.~\eq{APPROT} and~\eq{EXPLROT}, we find
\beq
e^{-i(\pi/2)S_x} S_z e^{i(\pi/2)S_x} = -S_y,\quad e^{-i(\pi/2)S_x}S_z = -S_y e^{-i(\pi/2)S_x}\,,
\eeq
so that we have
\beq
\ket{\psi_2'} = i S_y \ket{\psi_2}\,.
\eeq
The final result thus works out as
\bea
{\cal P}'(\theta) &=& \frac{d}{d\theta}\mele{\psi_2}{{\cal P}}{\psi_2} = \mele{\psi_2'}{{\cal P}}{\psi_2}
+ \mele{\psi_2}{{\cal P}}{\psi_2'}\nonumber\\
&=& -i \mele{\psi_2}{[S_y,{\cal P}]}{\psi_2}\,.
\eea

\begin{thebibliography}{28}
\bibitem [{NOT()}]{NOTEONEQUIV}%
  \BibitemOpen
  \href@noop {} {}\bibinfo {note} {Here we regard the double-well condensate
  and a two-component condensate (with an ``internal {J}osephson effect'') as
  interchangeable.}\BibitemShut {Stop}%
\end{thebibliography}
\end{document}